\documentclass[aps,preprint,prd,nofootinbib]{revtex4}
\usepackage{graphicx}
\usepackage{psfrag}

\begin{document}

\title{The weak decay $B_c$ to $Z(3930)$ and $X(4160)$ by Bethe-Salpeter method}
\author{Zhi-Hui Wang$^{[1],[2]}$\footnote{zhwang@nmu.edu.cn}, Yi Zhang$^{[1],[2]}$, Tianhong Wang$^{[3]}$, Yue Jiang$^{[3]}$, Guo-Li Wang$^{[4],[5]}$\\}
\address{$^1$ Key Laboratory of Physics and Photoelectric Information Functional Materials, North Minzu University, Yinchuan, 750021, China,\\
$^2$School of Electrical and Information Engineering, North Minzu University, Yinchuan, 750021, China,\\
$^3$School of Physics, Harbin Institute of
Technology, Harbin, 150001, China,\\
$^4$Department of Physics, Hebei University, Baoding, 071002, China,\\
$^5$Hebei Key Laboratory of High-precision Computation and
Application of Quantum Field Theory, Baoding, 071002, China}

 \baselineskip=20pt

\begin{abstract}
Considering $Z(3930)$ and $X(4160)$ as $\chi_{c2}(2P)$ and $\chi_{c2}(3P)$ states,
the semileptonic and nonleptonic of $B_c$ decays to $Z(3930)$ and $X(4160)$ are studied by
the improved Bethe-Salpeter(B-S) Method.
The form factors of decay are calculated through the overlap
integrals of the meson wave functions in the whole accessible kinematical range.
The influence of relativistic corrections are considered in the exclusive decays.
Branching ratios of $B_c$ weak decays to $Z(3930)$ and $X(4160)$ are predicted.
Some of the branching ratios are:
$Br(B_c^+\to Z(3930)e^+\nu_e)$$=(3.03^{+0.09}_{-0.16})\times 10^{-4}$ and $Br(B_c^+\to X(4160)e^+\nu_e)$$=(3.55^{+0.83}_{-0.35})\times 10^{-6}$.
These results may provide useful information to discover $Z(3930)$ and $X(4160)$ and the
necessary information for the phenomenological study of
$B_c$ physics.

 \vspace*{0.5cm}


\end{abstract}

\maketitle
\section{Introduction}
During the past decade years, more and more charmonium and charmonium-like states were discovered experimentally.
Such as the $X(3915)$ was reported by Belle Collaboration
in $\gamma\gamma\to\omega J/\psi$ process~\cite{3915}.
$Z(3930)$ was observed in the process $\gamma\gamma\to D\bar D$ by Belle Collaboration in 2006,
the corresponding mass and width were $M=3929\pm5\pm2$~MeV and
$\Gamma=29\pm10\pm2$ MeV, respectively~\cite{Z3930-Belle}.
In 2010, BABAR Collaboration also observed the  $Z(3930)$
in $\gamma\gamma$ production of the $D\bar D$ system, with the mass and width
being $M=3926.7\pm2.7\pm1.1$~MeV and
$\Gamma=21.3\pm3.8\pm3.6$ MeV, respectively~\cite{Z3930-BABAR}.
Now Particle Data Group(PDG) give lists the mass and
width of $Z(3930)$ as $M=3927.2\pm2.6$ MeV and $\Gamma=24\pm6$ MeV~\cite{PDG}.
And the properties of $Z(3930)$ are consistent with the expectations for the
$\chi_{c2}(2P)$ state~\cite{chic1,chic2,chic3}.
Then Belle Collaboration reported a new charmonium-like state $X(4160)$
from the processes $e^+e^-\to J/\psi D^*\bar D^*$,
which has the mass and width $M=(4156^{+25}_{-20}\pm15)$ MeV and $\Gamma=(139^{+111}_{-61}\pm21)$ MeV, respectively~\cite{39404160}.

The quark structures were still not fully understood in these charmonium-like states which are called XYZ states,
thus people studied the properties of XYZ states by different methods~\cite{bctoZ3930,Z39301,liux1,thwang,slzhu1,stephen1,kim,Lu,39401,zhuruilin,liuxiang2,41601,th6,zhaoqiang}.
In this work, we only consider two of them: $Z(3930)$ and $X(4160)$.
The structures of $Z(3930)$ and $X(4160)$ were already studied by some theoretical methods.
Ref.~\cite{bctoZ3930} studied $B_c$ semileptonic decay to $Z(3930)$ and $X(4160)$
which were assumed as $\chi_{c2}(2P)$ and $\chi_{c2}(3P)$ states.
According to study the vector-vector interaction within the framework of the hidden gauge formalism,
Ref.~\cite{Z39301} found that three resonances $Y(3940)$, $Z(3930)$ and $X(4160)$
which can be assigned to the states
with $J^{PC}=0^{++},~2^{++}$ and $2^{++}$, respectively.
Taking $Z(3915)$ and  $Z(3930)$ as $\chi_{c0}^\prime(2P)$ and $\chi_{c2}^\prime(2P)$, respectively.
Ref.~\cite{liux1} investigated the $X(3915)$ and $Z(3930)$ decays into $J/ \psi\omega$.
Ref.~\cite{thwang} studied the strong decay of $Z(3930)$ which was considered as  $\chi_{c2}^\prime(2P)$.
Ref.~\cite{slzhu1} studied the mass spectra of the hidden-charm
tetraquark states in the framework of QCD sum rules, and they got the $X(4160)$
may be classified as either the scalar or tensor $qc\bar q\bar c$ tetraquark state.
Using the NRQCD factorization approach, Ref.~\cite{zhuruilin} calculated the branching fractions of
$\Upsilon(nS)\to J/\psi+X$ with $X=X(3940)$ or $X=X(4160)$.
In Ref.~\cite{liuxiang2}, they also explored the properties and strong decays of $X(3940)$ and $X(4160)$
as the $\eta_c(3S)$ and $\eta_c(4S)$, respectively.
Ref.~\cite{41601} calculated the strong decay of $X(4160)$ which was assumed as $\chi_{c0}(3P)$,
$\chi_{c1}(3P)$, $\eta_{c2}(2D)$ or $\eta_c(4S)$ by the $^3P_0$ model.
Ref.~\cite{4S4160} studied the strong decays of $X(3940)$ and $X(4160)$ as the $\eta_c(3S)$ and $\eta_c(4S)$ with the $^3P_0$ model,
and the results showed that $\eta_c(4S)$ was not good candidate of $X(4160)$.
According to the mass spectra and the properties of $Z(3930)$ and $X(4160)$ in Ref.~\cite{bctoZ3930,Z39301,liux1,thwang,slzhu1},
$Z(3930)$ and $X(4160)$ have the possibility to be $\chi_{c2}(2P)$ and $\chi_{c2}(3P)$($J^{PC}=2^{++}$), respectively.

The interpretations of $Z(3930)$ and $X(4160)$ are not the major work in this paper,
we only consider $Z(3930)$ and $X(4160)$ as charmonium states
with the possible quantum numbers, then study their production in $B_c$ decays.
We will consider $Z(3930)$ and $X(4160)$
as $P-$wave charmonium states $\chi_{c2}(2P)$ and $\chi_{c2}(3P)$, respectively.
Then we focus on the productions of $Z(3930)$ and $X(4160)$
in exclusive weak decays of $B_c$ meson by the improved the Bethe-Salpeter(B-S) Method.
On the one hand, the $\chi_{c2}(2P)$ and $\chi_{c2}(3P)$ have larger relativistic correction than
that of $\chi_{c2}(1P)$, so a relativistic model is needed in a careful study;
on the other hand, this study can improve the knowledge of $B_c$ meson,
which is an ideal particle to study the weak decays, since it decays weakly only.
The properties of $B_c$ meson have been studied by different relativistic constituent quark models~\cite{bc1,bc2,bc3,bc4,bc5,Ebert11,Ebert12,Ivanov1,Ivanov2},
such as the covariant light-front quark model~\cite{bc6,bc7},
the perturbative QCD factorization approach~\cite{bc8} and so on.
We also have discussed the properties of $B_c$ meson by the improved B-S method,
include $B_c$ decays to $P-$wave mesons,
the rare weak decays and rare radiative decays of $B_c$, the nonleptonic charmless decays of $B_c$,
and so on~\cite{bc-pwave,heavy-light,bc9,bc10,bc11,bc12,bc13}.
In previous work, we only studied $B_c$ decays to $\chi_{c2}(1P)$ state~\cite{bc-pwave},
because when the final states are $\chi_{c2}(2P)$ and $\chi_{c2}(3P)$ states,
the corresponding branching ratios are very small,
and there were only limited data of $B_c$ available.
Now the Large Hadron Collider (LHC)
will produce as many as $5\times10^{10}$ $B_c$ events per year~\cite{lhc1,lhc2}.
The huge amount of $B_c$ events will provide us a chance
to study $B_c$ decay to $\chi_{c2}(2P)$ and $\chi_{c2}(3P)$ states,
and some channels also provide an opportunities to discover new particles in $B_c$ decays.

The paper is organized as follows.
In Sec.~II, we give the formulations of the exclusive semileptonic and
nonleptonic decays.
We show the hadronic weak-current matrix elements in Section.~III.
The wave functions of initial and final mesons are given in Sec.~IV.
The corresponding results and conclusions are presented in Sec.~V.
Finally in the Appendix, we present the instantaneous Bethe-Salpeter equation.

\section{The formulations of semileptonic decays and nonleptonic decays of $B_c$}
 In this section we present the formulations of
  semileptonic decays and nonleptonic decays of $B_c$ meson to $Z(3930)$ and $X(4160)$
which
are considered as $\chi_{c2}(2P)$ and $\chi_{c2}(3P)$ states, respectively.

\subsection{Semileptonic decays of $B_c$}

\begin{figure}
\centering
\includegraphics[height=5cm]{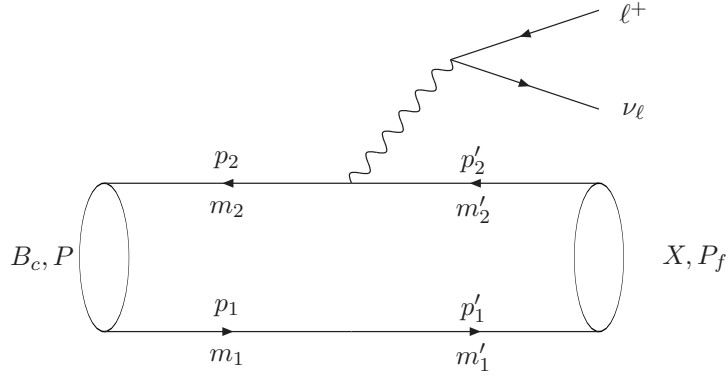}
\caption{\label{semileptonic}{Feynman diagram of the semileptonic decay $B_c\to X \ell^+\nu_\ell$,
where $X$ denotes $Z(3930)$ or $X(4160)$.}}
\end{figure}

The Feynman diagram of $B_c$ semileptonic decay to $Z(3930)$ or $X(4160)$
is shown in Fig. \ref{semileptonic}.
The corresponding amplitude for the decay can be written as
\begin{eqnarray}\label{T}
T=\frac{G_F}{\sqrt{2}}V_{bc}\bar{u}_{\nu_\ell}\gamma_{\mu}(1-\gamma_5)
v_{\ell}\langle X(P_f,\varepsilon)|J^{\mu}|B_c(P)\rangle\,,
\end{eqnarray}
where $V_{bc}$ is the CKM matrix element, $G_F$ is the Fermi constant,
$J^{\mu}=V^{\mu}-A^{\mu}$ is the charged weak current, $P$ and
$P_f$ are the momenta of the initial meson $B_c$ and the final
state, respectively. $\varepsilon$ is the polarization tensor for final meson.
The leptonic part $\bar{u}_{\nu_\ell}\gamma_{\mu}(1-\gamma_5)
v_{\ell}$ is model independent and easy to calculate.
 The hadronic part $\langle X(P_f,\varepsilon)|J^{\mu}|B_c(P)\rangle$ can be written
as,
\begin{eqnarray}\label{form}
&&\langle X(P_f,\varepsilon)|A^{\mu}|B_c(P)\rangle
=k(M+M_f){\varepsilon}^{\mu\alpha}\frac{P_{\alpha}}{M}
+{\varepsilon}_{\alpha\beta}\frac{P^{\alpha}P^{\beta}}{M^2}
(c_1P^{\mu}+c_2P_f^{\mu}),\nonumber\\
&& \langle X(P_f,\varepsilon)|V^{\mu}|B_c(P)\rangle=
\frac{2h}{M+M_f}i{\varepsilon}_{\alpha\beta}\frac{P^{\alpha}}{M}
{\epsilon}^{\mu\beta\rho\sigma}P_{\rho}{P_f}_{\sigma}\,.
\end{eqnarray}
where $k$, $c_1$, $c_2$, $h$ are the Lorentz invariant form factors,
$M$ is the mass of $B_c$, $M_f$ is the mass of the
charmonium in the final state.

In the case without considering polarization, we have the squared
decay-amplitude with the polarizations in final states being summed:
\begin{equation}
\Sigma_{s_\nu,s_l,S_{X}}|T|^2=\frac{G_F^2}{2}|V_{bc}|^2l_{\mu\nu}h^{\mu\nu}
\end{equation}
where $l_{\mu\nu}$ is the leptonic tensor:
$$l_{\mu\nu}=\Sigma_{s_\nu,s_l}\bar{{\upsilon}_l}(p_l){\gamma}_{\mu}(1-{\gamma}_5)
{u}_{{\nu}_l}(p_\nu)\bar{u}_{{\nu}_l}(p_\nu){\gamma}_{\nu}
(1-{\gamma}_5){\upsilon}_l(p_l),$$ and the hadronic tensor relating
to the weak-current in Eq.(\ref{T}) is
\begin{eqnarray}
&h^{\mu\nu} \equiv \Sigma_{S_{X}}\langle
B_c(P)|J^{\mu}|X(P_f)\rangle\langle
X(P_f)|J^{\nu}|B_c(P)\rangle\nonumber\\
& =-{\alpha}g^{\mu\nu}+{\beta}_{++}(P+P_f)^{\mu}(P+P_f)^{\nu}
+{\beta}_{+-}(P+P_f)^{\mu}(P-P_f)^{\nu}\nonumber\\
&
+{\beta}_{-+}(P-P_f)^{\mu}(P+P_f)^{\nu}+{\beta}_{--}(P-P_f)^{\mu}(P-P_f)^{\nu}\nonumber\\
&+i\gamma{\epsilon}^{\mu\nu\rho\sigma}(P+P_f)_{\rho}(P-P_f)_{\sigma},
\end{eqnarray}
where the functions $\alpha$, $\beta_{++}$, $\beta_{+-}$,
$\beta_{-+}$, $\beta_{--}$, $\gamma$ are related to the form factors.

The total decay width $\Gamma$ can be written as:
\begin{eqnarray}\label{total-g}
\Gamma&=&\frac{1}{2M(2\pi)^9}\int\frac{d^3\vec{P}_f}{2E_f}
\frac{d^3\vec{p}_l}{2E_l}\frac{d^3\vec{p}_{\nu}}{2E_{\nu}}
(2\pi)^4{\delta}^4(P-P_f-p_l-p_{\nu})\Sigma_{s_\nu,s_l,S_{X}}|T|^2,
\end{eqnarray}
where $E_f$, $E_l$ and $E_\nu$ are the energies of the charmonium,
the charged lepton and the neutrino respectively. If we define
$x\equiv E_l/M,\;\; y\equiv (P-P_f)^2/M^2$, the differential width
of the decay can be reduced to:
\begin{eqnarray}\label{differ}
&\displaystyle\frac{d^2\Gamma}{dxdy}=|V_{bc}|^2\frac{G_F^2M^5}{64{\pi}^3}\left\{
\frac{2\alpha}{M^2}(y-\frac{m_l^2}{M^2})\right.\nonumber\\
&\displaystyle+{\beta}_{++}\left[4\left(2x(1-\frac{M_f^2}{M^2}+y)-4x^2-y\right)
+\frac{m_l^2}{M^2}\left(8x+4\frac{M_f^2}{M^2}-3y-\frac{m_l^2}{M^2}\right)\right]\nonumber\\
&\displaystyle+({\beta}_{+-}+{\beta}_{-+})\frac{m_l^2}{M^2}
\left(2-4x+y-2\frac{M_f^2}{M^2}+\frac{m_l^2}{M^2}\right)
+{\beta}_{--}\frac{m_l^2}{M^2}\left(y-\frac{m_l^2}{M^2}\right)\nonumber\\
&\displaystyle\left.-\left[2{\gamma}y\left(1-\frac{M_f^2}{M^2}-4x+y+\frac{M_l^2}{M^2}\right)+
2\gamma\frac{M_l^2}{M^2}\left(1-\frac{M_f^2}{M^2}\right)\right]\right\}\,,
\end{eqnarray}

The total width of the decay is just
an integration of the differential width i.e. $\Gamma=\int dx\int
dy\frac{d^2\Gamma}{dxdy}$.

\subsection{Nonleptonic decays of $B_c$}

\begin{figure}
\centering
\includegraphics[height=5cm]{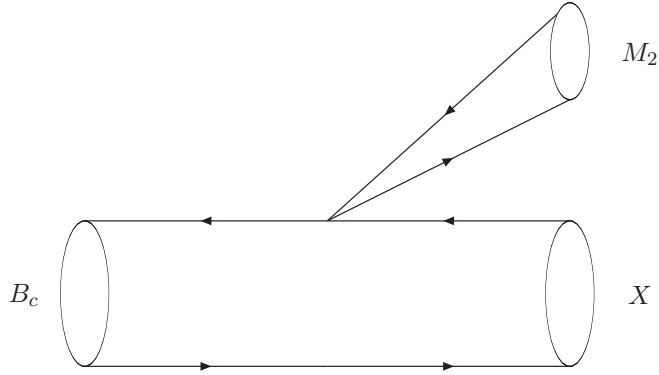}
\caption{\label{nonleptonic}{Feynman diagram of the nonleptonic decay $B_c\to X M_2$,
$X$ denote $Z(3930)$ or $X(4160)$, $M_2$} denote a light meson: $\pi, K, \rho,$ or $K^*$.}
\end{figure}

For the nonleptonic decay $B_c\to X+M_2$ in Fig.~\ref{nonleptonic},
the relevant effective Hamiltonian $H_{eff}$ is~\cite{Heff1,Heff2}:

\begin{eqnarray}\label{Heff}
H_{eff}=\frac{G_F}{\sqrt2}
\left\{V_{bc}[c_1(\mu)O_1^{bc}+c_2(\mu)O_2^{bc}]+h.c.\right\},
\end{eqnarray}
where $c_i(\mu)$ are the scale-dependent Wilson coefficients.
$O_i$ are the operators responsible for the decays constructed by four quark fields and
have the structure as follows:
\begin{eqnarray}\label{O1O2}
&&O_1^{bc}=[V_{ud}(\bar d_\alpha u_\alpha)_{V-A}+V_{us}(\bar s_\alpha u_\alpha)_{V-A}](\bar c_\beta b_\beta)_{V-A},
\nonumber\\
&&O_2^{bc}=[V_{ud}(\bar d_\alpha u_\beta)_{V-A}+V_{us}(\bar s_\alpha u_\beta)_{V-A}](\bar c_\beta b_\alpha)_{V-A},
\end{eqnarray}
where $(\bar q_1q_2)_{V-A}=\bar q_1\gamma^\mu(1-\gamma_5)q_2$.

Here we apply the so-called naive factorization to $H_{eff}$ \cite{naive},
the nonleptonic two-body decay amplitude $T$ can be reduced to a product of
a transition matrix element of a weak current $\langle X|J^\mu|B_c\rangle$
and an annihilation matrix element of another weak current $\langle M_2|J_\mu|0\rangle$:
\begin{eqnarray}\label{nonamplitude}
T=\langle XM_2|H_{eff}|B_c\rangle\approx
\frac{G_F}{\sqrt2}V_{bc}V_{ij}a_1\langle X|J^\mu|B_c\rangle\langle M_2|J_\mu|0\rangle,
\end{eqnarray}
$a_1=c_1+\frac{1}{N_c}c_2$ and $N_c=3$ is the number of colors.
The annihilation matrix element $\langle M_2|J_\mu|0\rangle$ is related to
the decay constant of $M_2$. When $M_2$ is a pseudoscalar meson~\cite{pseudo},
$$\langle M_2|J_\mu|0\rangle=if_{M_2}P_{M_2\mu},$$
where $f_{M_2}$ is the decay constant of meson $M_2$, and $P_{M_2}$ is the momentum of $M_2$.
When $M_2$ is a vector meson~\cite{vector},
$$\langle M_2|J_\mu|0\rangle=\epsilon_\mu f_{M_2}M_{M_2},$$
where $M_{M_2}$, $f_{M_2}$ and $\epsilon$ are the mass, decay constant and polarization vector of
the vector meson $M_2$, respectively. The decay constant of the meson can be obtained
either by theoretical model or by indirect experiment measurement.

In Eq.~(\ref{differ}) and Eq.~(\ref{nonamplitude}),
we find that the most important things to get the decay width of the corresponding decay
are to calculate hadronic weak-current matrix elements $\langle
X(P_f)|J^{\mu}|B_c(P)\rangle$.
We will give the detailed calculation of the hadronic weak-current matrix elements in the Section.~III.

\section{The hadronic weak-current matrix elements}

The calculation of the hadronic weak-current matrix element are different for different models.
In this paper, we combine the B-S method which is
based on relativistic B-S equation with Mandelstam formalism~\cite{Mand}
and relativistic wave functions to calculate the
hadronic matrix element.
The numerical values of wave functions
have been obtained by solving the full Salpeter equation which we
will introduce in Appendix. As an example, we consider the
semileptonic decay $B_c\to X\ell^+{\nu_\ell}$ in Fig.~\ref{semileptonic}. In this way,
at the leading order the hadronic matrix element can be written as
an overlap integral over the wave functions of initial and final mesons~\cite{BS1},
\begin{eqnarray}\label{matrix}
\langle X(P_f,\varepsilon)|J^{\mu}|B_c(P)\rangle=
\int\frac{d{\vec{q}}}{(2\pi)^3}{\rm Tr}\left[
\bar{\varphi}^{++}_{_{P_{f}}}(\vec {q}_{_f})\frac{\not\!P}{M}
{\varphi}^{++}_{_P}({\vec{q}})\gamma^{\mu}(1-\gamma_5)\right]\, ,
\end{eqnarray}
where $\vec{q}$ ($\vec{q}_{_f}$) is the relative three-momentum
between the quark and anti-quark in the initial (final) meson and
$\vec{q}_{_f}=\vec{q}-\frac{m'_1}{m'_1+m'_2}{\vec{P_f}}$.
${\vec{P_f}}$ is the three dimensional
momentum of $X$, ${\varphi}^{++}_P(\vec q)$ is the positive
Salpeter wave function of $B_c$ meson and
${\varphi}^{++}_{P_f}(\vec q_f)$
is the positive Salpeter wave function of $X$ meson,
$\bar{\varphi}^{++}_{_{P_f}}=\gamma_0({\varphi}^{++}_{_{P_f}})^{\dagger}\gamma_0$.
The detailed calculation of the hadronic matrix element Eq.~(\ref{matrix})
which is a function of final meson momentum $P_f$ were discussed by Ref.~\cite{BS1},
so the Eq.~(\ref{matrix}) is suitable for the whole kinetic region.
We have calculated $B_c$ weak decays to $S-$wave and $P-$wave mesons~\cite{bc-pwave,heavy-light,bc13}
with this hadronic matrix element in previous work,
and the results were consistent with the results of some other different models.
So the $B_c$ weak decays to $Z(3930)$ and $X(4160)$ are calculated by the same metnod in this work.
The corresponding Salpeter wave functions for the different mesons are shown in the next section.

\section{The Relativistic Wave functions of Meson}

\subsection{ For $B_c$ meson with quantum number
$J^{P}=0^{-}$}

The general form for the relativistic wave function of
pseudoscalar meson $B_c$ can be written as~\cite{w1}:
\begin{eqnarray}\label{aa01}
\varphi_{0^-}(\vec q)&=&\Big[f_1(\vec q){\not\!P}+f_2(\vec q)M+
f_3(\vec q)\not\!{q_\bot}+f_4(\vec q)\frac{{\not\!P}\not\!{q_\bot}}{M}\Big]\gamma_5,
\end{eqnarray}
where $M$ is the mass of the pseudoscalar meson, and
$f_i(\vec q)$ are functions of $|\vec q|^2$. Due to
the last two equations of Eq.~(\ref{eq11}):
$\varphi_{0^-}^{+-}=\varphi_{0^-}^{-+}=0$, we have:
\begin{eqnarray}\label{constrain}
f_3(\vec q)&=&\frac{f_2(\vec q)
M(-\omega_1+\omega_2)}{m_2\omega_1+m_1\omega_2},~~~
f_4(\vec q)=-\frac{f_1(\vec q)
M(\omega_1+\omega_2)}{m_2\omega_1+m_1\omega_2}.
\end{eqnarray}
where $m_1, m_2$ and
$\omega_1=\sqrt{m_1^{2}+\vec{q}^2},\omega_2=\sqrt{m_2^{2}+\vec{q}^2}$ are
the masses and the energies of
 quark and anti-quark in $B_c$ mesons, $q_{_\bot}=q-(q\cdot P/M^2)P$, and $q_{\bot}^2=-|\vec q|^2$.

The numerical values of radial wave functions $f_1$, $f_2$ and
eigenvalue $M$ can be obtained by solving the first two Salpeter equations in
 Eq.~(\ref{eq11}).
According to the Eq.~(\ref{eq10}) the relativistic positive wave function
of pseudoscalar meson $B_c$ in C.M.S can be written as \cite{w1}:
\begin{eqnarray}\label{0-postive}
{\varphi}^{++}_{0^-}(\vec{q})=b_1
\left[b_2+\frac{\not\!{P}}{M}+b_3\not\!{q_{\bot}}
+b_4\frac{\not\!{q_{\bot}}\not\!{P}}{M}\right]{\gamma}_5,
\end{eqnarray}
where the $b_i$s ($i=1,~2,~3,~4$) are related to the original
radial wave functions $f_1$, $f_2$, quark masses $m_1$, $m_2$, quark energy $w_1$, $w_2$,
and meson mass $M$:
$$b_1=\frac{M}{2}\left({f}_{1}(\vec{q})
+{f}_{2}(\vec{q})\frac{m_1+m_2}{\omega_1+\omega_2}\right),
b_2=\frac{\omega_1+\omega_2}{m_1+m_2}, b_3=-\frac{(m_1-m_2)}{m_1\omega_2+m_2\omega_1},
b_4=\frac{(\omega_1+\omega_2)}{(m_1\omega_2+m_2\omega_1)}.$$

\subsection{For $Z(3930)$ and $X(4160)$ mesons with quantum number
$J^{P}=2^{++}$}

Considering $Z(3930)$ and $X(4160)$ as $\chi_{c2}(2P)$ and $\chi_{c2}(3P)$,
the general expression of the relativistic wave function can
be written as~\cite{mass1}

\begin{eqnarray}
&\displaystyle {\varphi}_{2^{++}}(\vec{q}_f)={\varepsilon}_{\mu
\nu}q^{\nu}_{f\bot}\{q^{\mu}_{f\bot}[f'_1(\vec{q}_f)+\frac{\not\!P_f}{M_f}f'_2(\vec{q}_f)
+\frac{\not\!q_{f\bot}}{M_f}f'_3(\vec{q}_{f\bot})
+\frac{\not\!P_f\not\!q_{f\bot}}{M_f^2}f'_4(\vec{q}_f)]\nonumber\\
&\displaystyle
+{\gamma}^{\mu}[M_ff'_5(\vec{q}_f)+\not\!P_ff'_6(\vec{q}_f)+\not\!q_{f\bot} f'_7(\vec{q}_f)]+
\frac{i}{M_f}f'_8(\vec{q}_f){\epsilon}^{\mu\alpha\beta\delta}
P_{f\alpha}q_{f\bot\beta}{\gamma}_{\delta}{\gamma}_5\}\,,
\end{eqnarray}

with the constraint on the components of the wave function:
$$f'_1(\vec{q}_f)=\frac{[q_{f\bot}^{2}f'_3(\vec{q}_f)+M_f^2f'_5(\vec{q}_f)]}{M_fm'_1}\,,\ \
f'_2(\vec{q}_f)=0\,,\;\;\; f'_7(\vec{q}_f)=0\,,\;\;\;
f'_8=\frac{f'_6(\vec{q}_f)M_f}{m'_1}.$$
Then we have the reduced wave function
${\varphi}_{2^{++}}(\vec{q}_f)$ as:
\begin{eqnarray}
&\displaystyle{\varphi}^{++}_{\chi_{c2}}(\vec{q}_f)={\varepsilon}_{\mu
\nu}q^{\nu}_{f\bot}\{q^{\mu}_{f\bot}[a_1+a_2\frac{\not\!P_f}{M_f}
+a_3\frac{{\not\!q_{f\bot}}}{M_f}\nonumber\\
&\displaystyle +a_4\frac{\not\!q_{f\bot}\not\!P_f}{M_f^2}]
+{\gamma}^{\mu}[a_5+a_6\frac{\not\!P_f}{M_f}+a_7\frac{\not\!q_{f\bot}}{M_f}+
a_8\frac{\not\!P_f\not\!q_{f\bot}}{M_f^2}]\}\,,
\end{eqnarray}
with
\begin{eqnarray}
&\displaystyle a_1=\frac{q^2_{f\bot}}{2M_fm'_1}n_1
+\frac{(f'_5(\vec{q}_f)w'_2-f'_6(\vec{q}_f)m'_2)M_f}{2m'_1w'_2}\,,\ \
\displaystyle a_2=\frac{(f'_6(\vec{q}_f)w'_2-f'_5(\vec{q}_f)m'_2)M_f}{2m'_1w'_2}\,,\nonumber\\
&\displaystyle
a_3=\frac{1}{2}n_1+\frac{f'_6(\vec{q}_f)M_f^2}{2m'_1w'_2}\,,\;\;\;\;
a_4=\frac{1}{2}(-\frac{w'_1}{m'_1})n_1+\frac{f'_5(\vec{q}_f)M_f^2}{2m'_1w'_2}\,,\nonumber\\
&\displaystyle
a_5=\frac{M_f}{2}n_2,a_6=\frac{M_fm'_1}{2w'_1}n_2\,,\;\;
a_7=0\,,\;\;
a_8=\frac{M_f^2}{2w'_1}n_2\,,\nonumber\\
&\displaystyle
n_1=\frac{1}{2}(f'_3(\vec{q}_f)+f'_4(\vec{q}_f)\frac{m'_1}{w'_1})\,,\;\;\;
n_2=\frac{1}{2}(f'_5(\vec{q}_f)-f'_6(\vec{q}_f)\frac{w'_1}{m'_1}),.\nonumber
\end{eqnarray}

\begin{figure}[htbp]
\centering
\includegraphics[height=5cm]{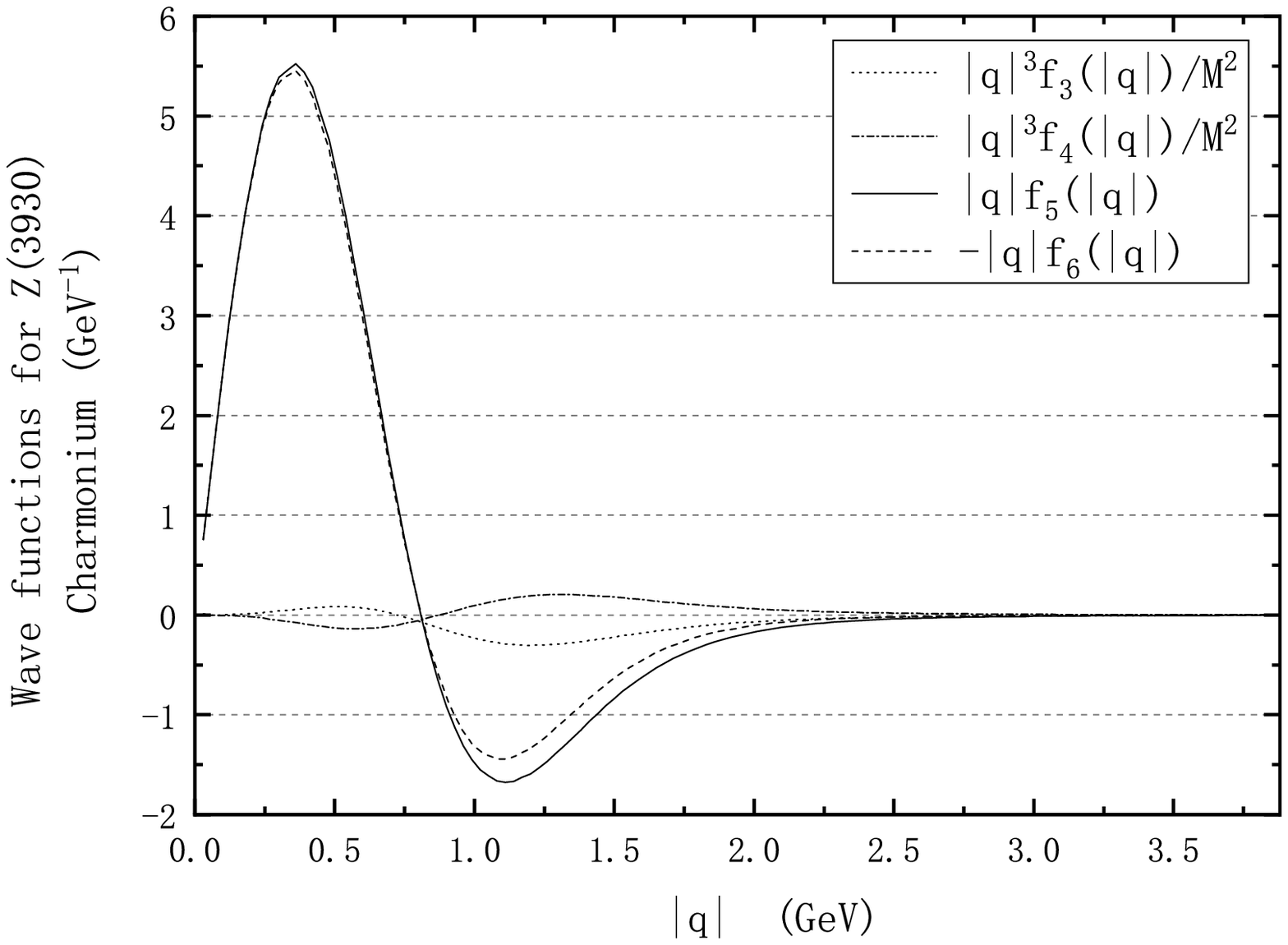}
\includegraphics[height=5cm]{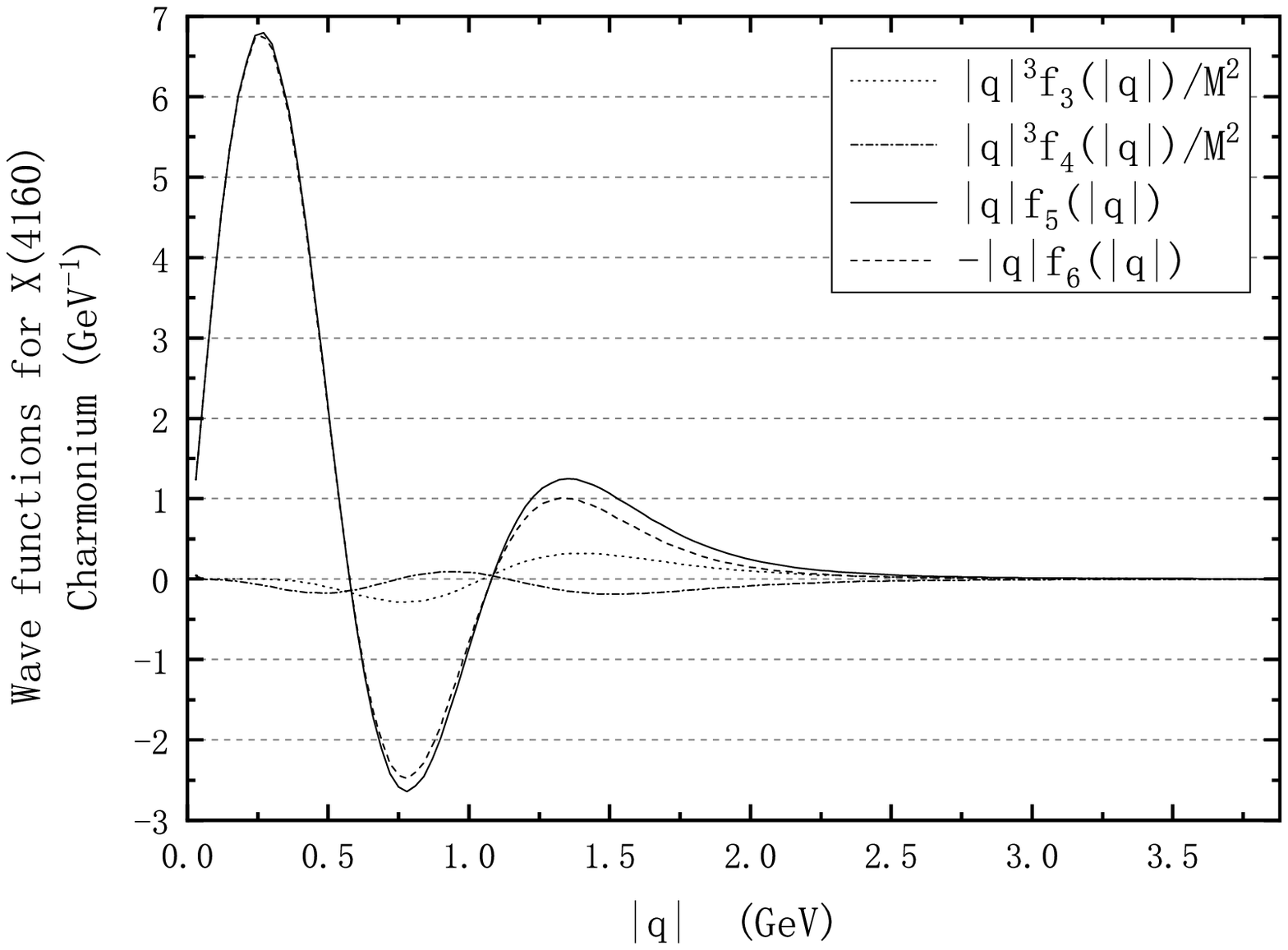}
\caption{\label{wavefunction}The wave functions of $Z(3930)$ and $X(4160)$.}
\end{figure}

Where $M_f$, $P_f$, $f'_i(\vec q_f)$ are the mass,
momentum and the radial wave functions of $Z(3930)$ and $X(4160)$, respectively.
$m'_1, m'_2$ and
$\omega'_1=\sqrt{m_1^{\prime2}+\vec{q}_f^2},\omega'_2=\sqrt{m_2^{\prime2}+\vec{q}_f^2}$ are
the masses and the energies of
 quark and anti-quark in $Z(3930)$ and $X(4160)$.
To show the numerical
results of wave functions explicitly, we plot the wave functions
of $Z(3930)$ and $X(4160)$ states in
Fig.~\ref{wavefunction}.

\section{Number results and discussions}

In order to fix Cornell potential in Eq.(\ref{eq16}) and masses of quarks,
 we take these parameters: $a=e=2.7183,
\lambda=0.21$ GeV$^2$, ${\Lambda}_{QCD}=0.27$ GeV, $\alpha=0.06$
GeV, $m_b=4.96$ GeV, $m_c=1.62$ GeV, $etc$~\cite{mass1},
 which are best to fit the mass spectra of $B_c$ and other heavy meson states.
Taking these parameters to B-S equation,
and solving the B-S equation numerically,
we get the masses of $Z(3930)$, $X(4160)$ and $B_c$ as:
$M_{Z(3930)}=(3.926\pm0.167)$ GeV, $M_{X(4160)}=(4.156\pm0.170)$ GeV, $M_{B_c}=(6.276\pm0.303)$ GeV,
varying all the
input parameters ($\lambda$, ${\Lambda}_{QCD}$, $\alpha$, $etc$) simultaneously within $\pm5\%$ of the central values,
we also obtain the uncertainties of masses,
and the corresponding wave functions were obtained in Section.IV.
Then we can calculate the semileptonic decays and nonleptonic decays
of $B_c$ to $Z(3930)$ and $X(4160)$.

\subsection{The semileptonic decays}

\begin{figure}[htbp]
\centering
\includegraphics[height=5cm]{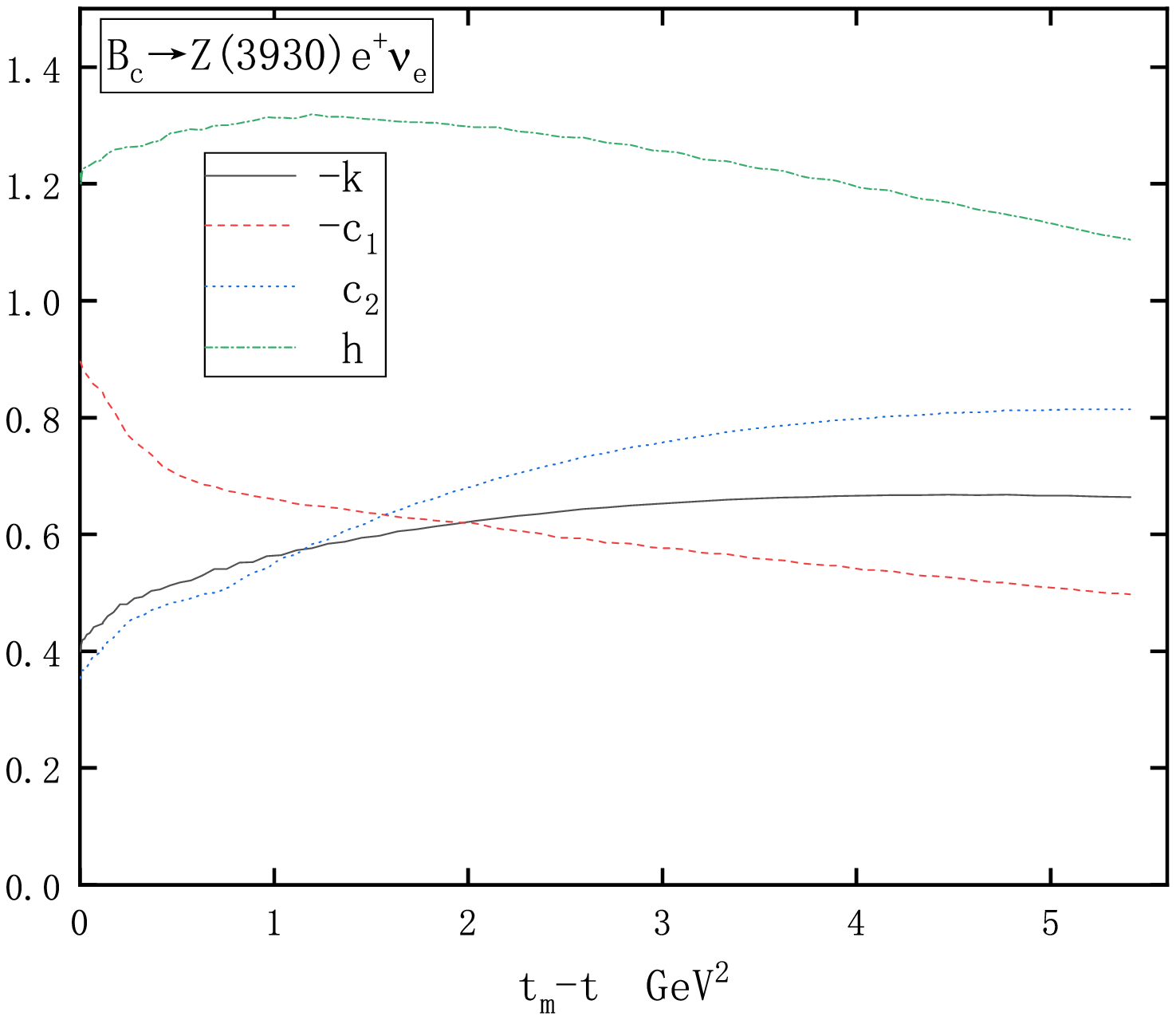}
\includegraphics[height=5cm]{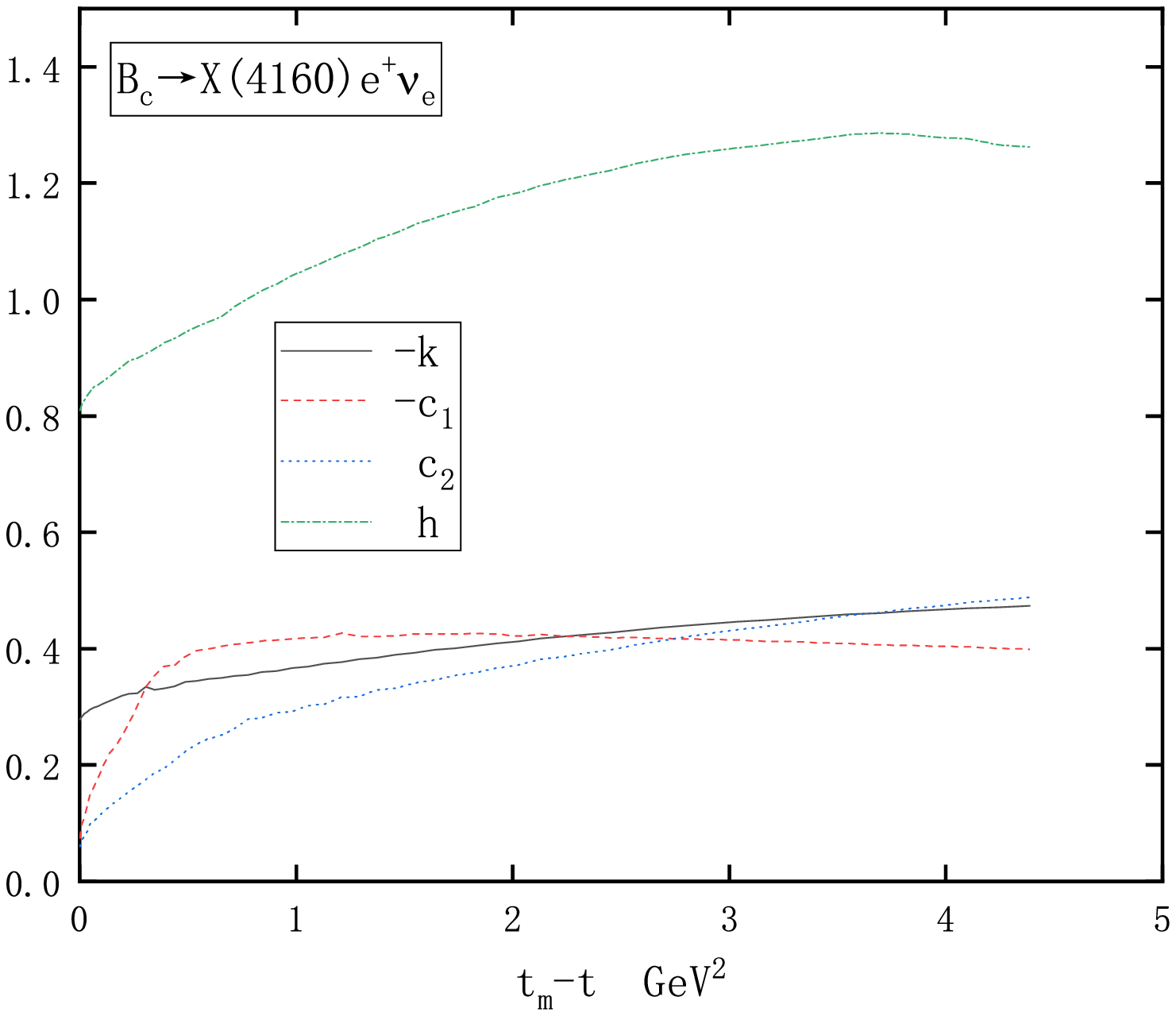}
\caption{\label{formfactor}The form factor of semileptonic decay $B_c$ to $Z(3930)$ and $X(4160)$.}
\end{figure}
\begin{figure}[htbp]
\centering
\includegraphics[height=5cm]{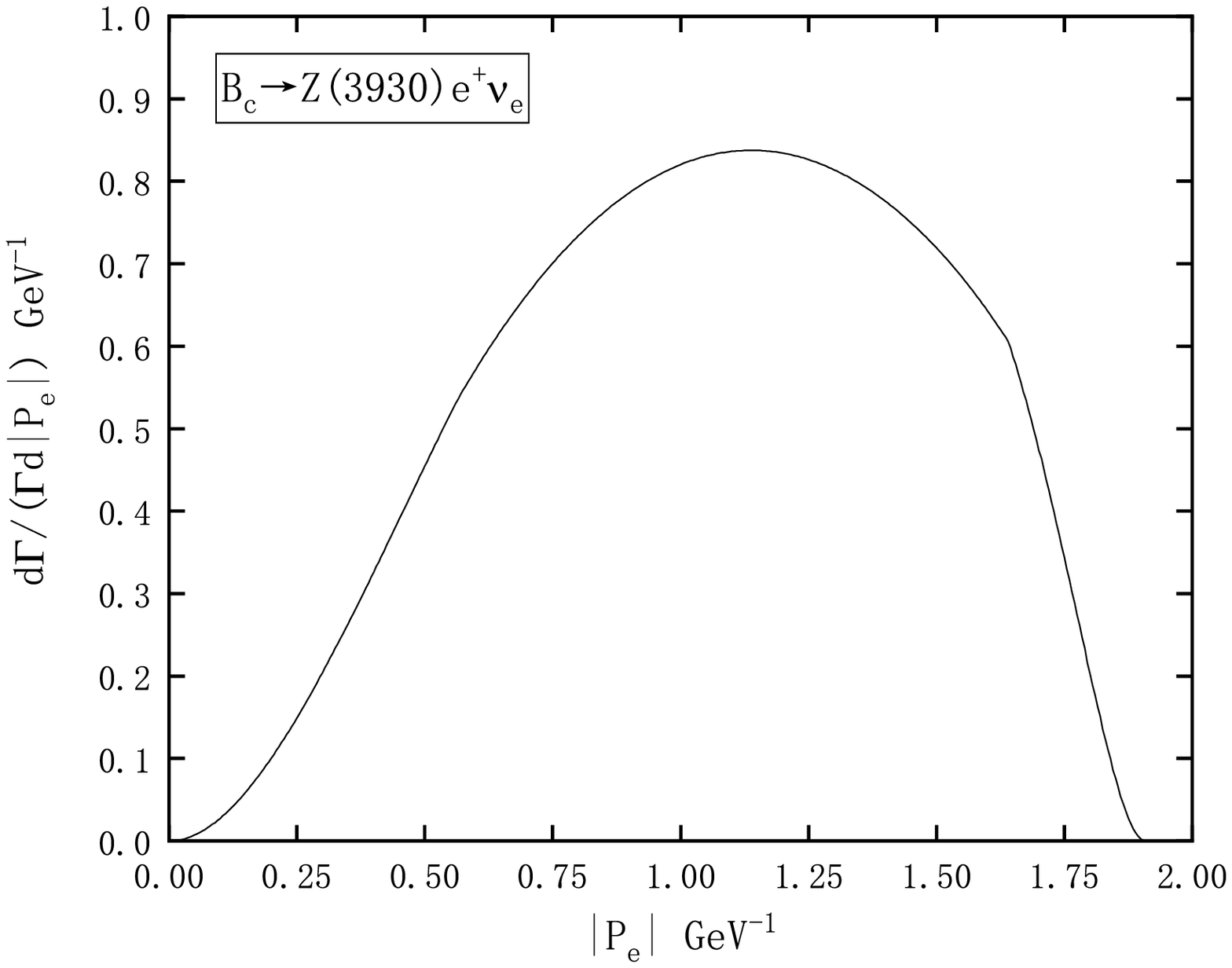}
\includegraphics[height=5cm]{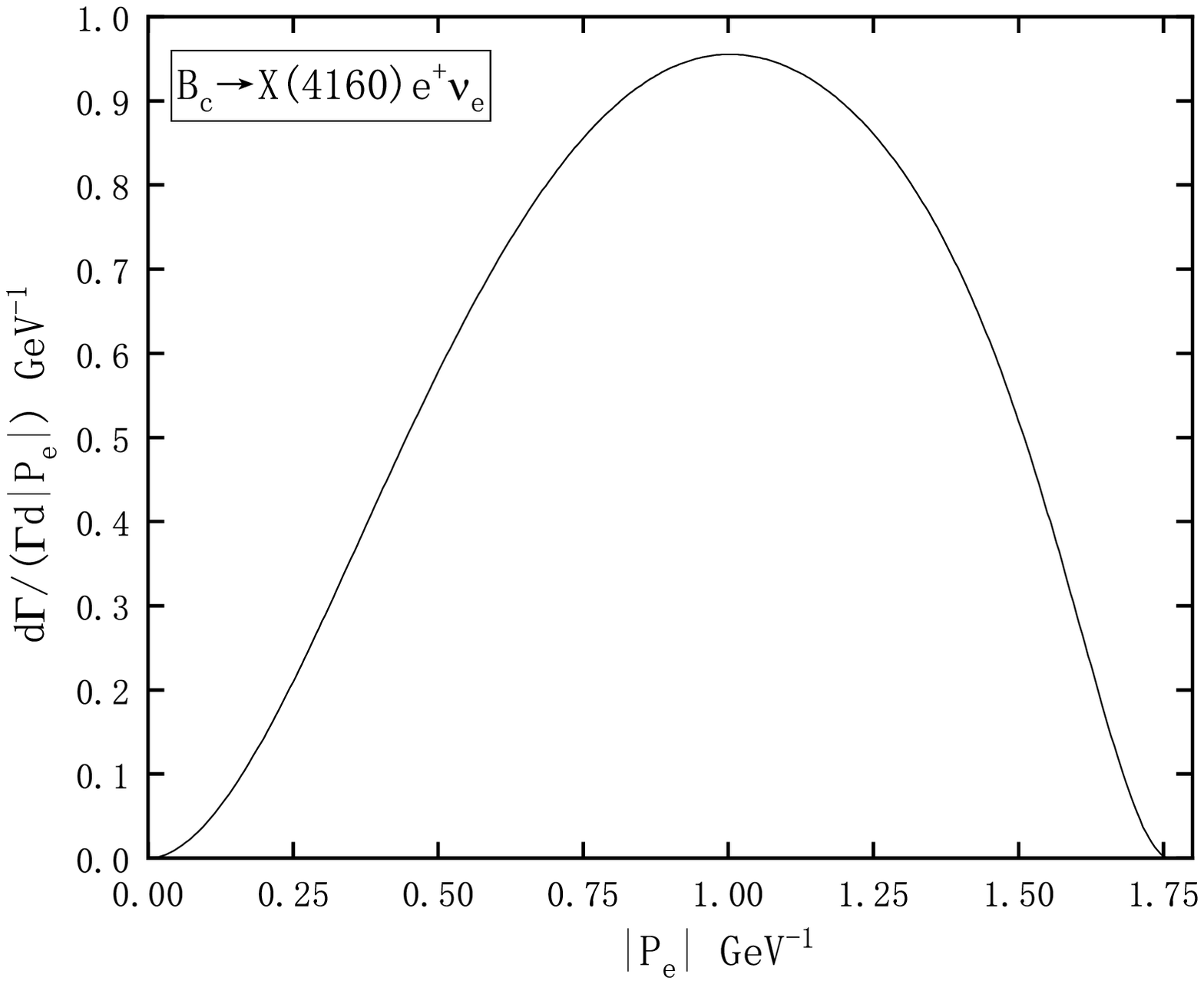}
\caption{\label{energyspectra}The leptonic energy spectra of semileptonic decay $B_c$ to $Z(3930)$ and $X(4160)$.}
\end{figure}
In order to calculate the semileptonic decays of $B_c$ to $Z(3930)$ and $X(4160)$,
we use the central values of the CKM matrix elements:
$V_{cb}=0.0406$,
and other constants: $G_F=1.166\times10^{-5}$ GeV$^{-2}$,
which are taken from PDG~\cite{PDG}.
Taking the masses and the corresponding wave functions to Eq.~(\ref{matrix}),
we represent the hadronic transition weak-current matrix elements as
proper integrations of the components of the B-S wave functions.
And the hadronic weak-current matrix element can be written as the form factors
$k$, $c_1$, $c_2$, $h$.
The form factors are related to four-momentum transfer squared $t=(P-P_f)^2=M^2+M_f^2-2ME_f$
which provides the kinematic range for the semileptonic decay of $B_c$.
It varies from $t=0$ to $t=5.52$ GeV$^2$ for the decays to $Z(3930)$ and
from $t=0$ to $t=4.48$ GeV$^2$ for the decays to $X(4160)$.
In Fig.~\ref{formfactor} we give the relations of $(t_m-t)$($t_m=(M-M_f)^2$ is the maximum of $t$) and the form factors.
Taking the form factor to the Eq.~(\ref{differ}),
then we will get the leptonic energy spectra $\frac{d\Gamma}{\Gamma dP_e}$ for semileptonic $B_c$
decay to $Z(3930)$ and $X(4160)$,
the leptonic energy spectra are plotted in Fig.~\ref{energyspectra}
which are related to the momentum of the final mesons.

Using the leptonic energy spectra,
we calculate the decay widths of the semileptonic
$B_c \to X\ell^+ \nu_\ell$ ($X=Z(3930)$ or $X(4160)$, $\ell=e,\mu,\tau$)
and give the results in Table.~\ref{semidecaywidth}.
Since $m_{\tau}$ is very large and $m_e\simeq m_\mu$
is quite a good approximation for the $B_c$ meson
decays, thus only the cases where the lepton is an electron or $\tau$ are given in Table.~\ref{semidecaywidth}.
Because of the larger kinematic ranges and the different wave functions in Fig.~\ref{wavefunction},
the corresponding decay widths of $B_c^+\to Z(3930)$ are larger than these of $B_c^+\to X(4160)$.

\begin{table}\caption{\label{semidecaywidth}The decay widths of exclusive semileptonic decays of $B_c$ to $Z(3930),X(4160)$ (in $10^{-15}$GeV).}
\begin{center}
\begin{tabular}{|c|c|}
\hline \hline
Mode&Ours  \\
\hline
 $B_c^+\to$ $Z(3930)$$e^+\bar\nu_e$&$(4.39_{-0.24}^{+0.13})\times 10^{-1}$\\
$B_c^+\to$ $Z(3930)$$\tau^+\bar\nu_\tau$&$(0.78_{-0.42}^{+0.31})\times 10^{-3}$\\
\hline
 $B_c^+\to$$X(4160)$$e^+\bar\nu_e$&$(5.14_{-0.49}^{+0.83})\times10^{-3}$\\
$B_c^+\to$$X(4160)$$\tau^+\bar\nu_\tau$&$(3.80_{-0.38}^{+0.45})\times10^{-6}$\\
\hline \hline
\end{tabular}
\end{center}
\end{table}

\subsection{The nonleptonic decays}

We only consider two-body nonleptonic decays of $B_c^+$ to $Z(3930)$ and $X(4160)$,
and another meson is light meson.
Thus, the hadronic transition
matrix elements of weak currents have a fixed momentum transfer.
To calculate the decay widths basis on Eq.~\ref{nonamplitude},
we only need to calculate the annihilation matrix element $\langle M_2|J_\mu|0\rangle$ which is related to
the decay constant of $M_2$.
The masses and decay constants are:
$M_\pi=0.140$ GeV,      $f_\pi=0.130$ GeV,
$M_\rho=0.775$ GeV,      $f_\rho=0.205$ GeV,
$M_K=0.494$ GeV,      $f_K=0.156$ GeV,
$M_{K^*}=0.892$ GeV,      $f_{K^*}=0.217$ GeV~\cite{PDG,kk}, respectively.
And the corresponding CKM matrix elements are:
$V_{ud}=0.974$ and $V_{us}=0.2252$.
Using the form factors of $B_c$ nonleptonic decays and the decay corresponding constants,
we show the nonleptonic decay widths which are related to the parameter $a_1$ in Table.~\ref{nondecaywidth}.
The results of $B_c$ nonleptonic decay are affected by the CKM matrix elements, so the results of light mesons $\pi, \rho$
are larger than the ones of light mesons $K, K^*$ in Table.~\ref{nondecaywidth}, respectively.

\begin{table}\caption{\label{nondecaywidth}The decay widths of exclusive nonleptonic decays of $B_c$ to $Z(3930),X(4160)$ (in $10^{-15}$GeV).}
\begin{center}
\begin{tabular}{|c|c|}
\hline \hline
Mode&Ours  \\
\hline
 $B_c^+\to$ $Z(3930)$+$\pi$&$(1.88_{-0.66}^{+0.49})\times 10^{-3}a^2_1$\\
$B_c^+\to$$Z(3930)$+$K$&$(1.38_{-0.51}^{+0.37})\times 10^{-4}a^2_1$\\
$B_c^+\to$$Z(3930)$+$\rho$&$(6.26_{-1.48}^{+1.42})\times 10^{-3}a^2_1$\\
$B_c^+\to$$Z(3930)$+$K^*$&$(3.82_{-0.86}^{+0.61})\times 10^{-4}a^2_1$\\
\hline
$B_c^+\to$$X(4160)$+$\pi$&$(6.89_{-1.15}^{+1.50})\times 10^{-5}a^2_1$\\
$B_c^+\to$$X(4160)$+$K$&$(4.71_{-0.79}^{+0.82})\times 10^{-6}a^2_1$\\
$B_c^+\to$$X(4160)$+$\rho$&$(2.37_{-0.42}^{+0.56})\times 10^{-3}a^2_1$\\
$B_c^+\to$$X(4160)$+$K^*$&$(1.64_{-0.33}^{+0.45})\times 10^{-4}a^2_1$\\
\hline \hline
\end{tabular}
\end{center}
\end{table}

\begin{table}\caption{\label{Branch}The branching ratio(in $\%$) of exclusive semileptonic decay $B_c$ to $Z(3930),X(4160)$ with the lifetime of $B_c$:$\tau_{B_c}=0.453$ps.}
\begin{center}
\begin{tabular}{|c|c|c|c|}
\hline \hline
Mode&Results&Mode&Results  \\
\hline
 $B_c^+\to$ $Z(3930)$$e^+\bar\nu_e$&$(3.03_{-0.16}^{+0.09})\times 10^{-2}$&$B_c^+\to$$X(4160)$$e^+\bar\nu_e$&$(3.55_{-0.35}^{+0.83})\times 10^{-4}$\\
$B_c^+\to$$Z(3930)$$\tau^+\bar\nu_\tau$&$(0.55_{-0.30}^{+0.22})\times 10^{-4}$&$B_c^+\to$$X(4160)$$\tau^+\bar\nu_\tau$&$(2.62_{-0.26}^{+0.31})\times 10^{-7}$\\
$B_c^+\to$$Z(3930)$+$\pi$&$(1.68_{-0.58}^{+0.44})\times 10^{-4}$&$B_c^+\to$$X(4160)$+$\pi$&$(6.17_{-1.02}^{+1.35})\times 10^{-6}$\\
$B_c^+\to$$Z(3930)$+$K$&$(1.24_{-0.46}^{+0.33})\times 10^{-5}$&$B_c^+\to$$X(4160)$+$K$&$(4.21_{-0.70}^{+0.75})\times 10^{-7}$\\
$B_c^+\to$$Z(3930)$+$\rho$&$(5.61_{-1.33}^{+1.28})\times 10^{-4}$&$B_c^+\to$$X(4160)$+$\rho$&$(2.12_{-0.37}^{+0.42})\times 10^{-4}$\\
$B_c^+\to$$Z(3930)$+$K^*$&$(3.43_{-0.78}^{+0.54})\times 10^{-5}$&$B_c^+\to$$X(4160)$+$K^*$&$(1.47_{-0.29}^{+0.41})\times 10^{-5}$\\
\hline \hline
\end{tabular}
\end{center}
\end{table}

In order to compare the numerical values with experimental measurements in the future,
Taking the values $a_1=1.14$ for nonleptonic decays~\cite{Heff1,Heff2},
combining the life time of $B_c$ meson, we calculate the branching ratios
of the decays and list them in Table.~\ref{Branch}.
Because of $B_c\to$$Z(3930)$, $X(4160)$ have small kinematic ranges and the wave functions have some minus parts in $Z(3930)$, and $X(4160)$,
comparing our results with  $B_c$ decays to $\chi_{c2}(1P)$ in Ref.~\cite{bc-pwave},
the results are smaller than the results of $B_c$ decay to $\chi_{c2}(1P)$.
The uncertainties of decay widths and branching ratios shown in Table.~\ref{semidecaywidth},
Table.~\ref{nondecaywidth} and Table.~\ref{Branch}, which are
very large. The large uncertainties not only come from the
phase spaces, but also from the variation of the node of the $2P$ and $3P$ wave functions,
which means that a small change of node location will result in large uncertainties.

In summary, considering $Z(3930)$ and $X(4160)$ as $\chi_{c2}(2P)$ and $\chi_{c2}(3P)$ states, respectively,
we study the semileptonic and nonleptonic $B_c$ decays to $Z(3930)$ and $X(4160)$ by
the improved B-S method which consider the relativistic correction.
According to the Mandelstam formalism and the relativistic wave functions of heavy mesons,
we get the corresponding decay form factors,
and obtain the corresponding
decay widths and branching ratios.
Because of the minus value in the wave functions of $Z(3930)$ and $X(4160)$ and the small CKM $V_{bc}$,
the decay widths and branching ratios are very small.
But now the Large Hadron Collider (LHC)
will produce as many as $5\times10^{10}$ $B_c$ events per year~\cite{lhc1,lhc2}.
If sufficient events can be observed, some channels will provide us a sizable ratios,
such as the branching ratios of the order of ($10^{-6}$) could be measured
precisely at the LHC,
and maybe they will detect the productions of $Z(3930)$ and $X(4160)$ in $B_c$ exclusive
weak semileptonic and nonleptonic decay.
Then our results will provide a new way to observe the $Z(3930)$ and $X(4160)$ and the
necessary information for the study of
$B_c$ meson.

 \noindent
{\Large \bf Acknowledgements}
This work was supported in part by
the National Natural Science Foundation of China (NSFC) under
Grant No.~11865001 and No.~11575048,
the Natural Science Foundation of Ningxia(2019AAC03127),
the CAS "Light of West China" Program and the Third Batch of Ningxia Youth Talents Supporting Program.

\appendix{
\section{Instantaneous Bethe-Salpeter Equation}

In this section, we briefly review the Bethe-Salpeter(B-S) equation and
its instantaneous one, the Salpeter equation.

The B-S equation is read as~\cite{BS}:
\begin{equation}
(\not\!{p_{1}}-m_{1})\chi(q)(\not\!{p_{2}}+m_{2})=
i\int\frac{d^{4}k}{(2\pi)^{4}}V(P,k,q)\chi(k)\;, \label{eq1}
\end{equation}
where $\chi(q)$ is the B-S wave function, $V(P,k,q)$ is the
interaction kernel between the quark and antiquark, and $p_{1},
p_{2}$ are the momentum of the quark 1 and anti-quark 2.

We divide the relative momentum $q$ into two parts,
$q_{\parallel}$ and $q_{\perp}$,
$$q^{\mu}=q^{\mu}_{\parallel}+q^{\mu}_{\perp}\;,$$
$$q^{\mu}_{\parallel}\equiv (P\cdot q/M^{2})P^{\mu}\;,\;\;\;
q^{\mu}_{\perp}\equiv q^{\mu}-q^{\mu}_{\parallel}\;.$$

B-S equation Eq.~(\ref{eq1}) is a four dimension covariant equation,
in order to solve the Eq.~(\ref{eq1}),
we will take the instantaneous approximation in the interaction kernel $V(P,k,q)$,
then the B-S equation will lose the covariance.
The effect of instantaneous approximation in $V(P,k,q)$ could be corrected by the
retardation effects in $V(P,k,q)$.
But the retardation effects in $V(P,k,q)$ are very small for the heavy mesons~\cite{eff1,eff2,eff3},
this means that the influence of the instantaneous approximation on the covariance of B-S equation
are very small for the heavy mesons.
The instantaneous approximation in $V(P,k,q)$ almost
don't influence the wave functions,
and the decay matrix elements which involve the heavy mesons mostly unchanged~\cite{eff1}.
Our model mostly keeps the covariance in the calculation, and the weak decay results also satisfy the Lorentz-covariance.

In instantaneous approach, the kernel $V(P,k,q)$ takes the simple
form~\cite{Salp}:
$$V(P,k,q) \Rightarrow V(|\vec k-\vec q|)\;.$$

Let us introduce the notations $\varphi_{p}(q^{\mu}_{\perp})$ and
$\eta(q^{\mu}_{\perp})$ for three dimensional wave function as
follows:
$$
\varphi_{p}(q^{\mu}_{\perp})\equiv i\int
\frac{dq_{p}}{2\pi}\chi(q^{\mu}_{\parallel},q^{\mu}_{\perp})\;,
$$
\begin{equation}
\eta(q^{\mu}_{\perp})\equiv\int\frac{dk_{\perp}}{(2\pi)^{3}}
V(k_{\perp},q_{\perp})\varphi_{p}(k^{\mu}_{\perp})\;. \label{eq5}
\end{equation}
Then the BS equation can be rewritten as:
\begin{equation}
\chi(q_{\parallel},q_{\perp})=S_{1}(p_{1})\eta(q_{\perp})S_{2}(p_{2})\;.
\label{eq6}
\end{equation}
The propagators of the two constituents can be decomposed as:
\begin{equation}
S_{i}(p_{i})=\frac{\Lambda^{+}_{ip}(q_{\perp})}{J(i)q_{p}
+\alpha_{i}M-\omega_{i}+i\epsilon}+
\frac{\Lambda^{-}_{ip}(q_{\perp})}{J(i)q_{p}+\alpha_{i}M+\omega_{i}-i\epsilon}\;,
\label{eq7}
\end{equation}
with
\begin{equation}
\omega_{i}=\sqrt{m_{i}^{2}+q^{2}_{_T}}\;,\;\;\;
\Lambda^{\pm}_{ip}(q_{\perp})= \frac{1}{2\omega_{ip}}\left[
\frac{\not\!{P}}{M}\omega_{i}\pm
J(i)(m_{i}+{\not\!q}_{\perp})\right]\;, \label{eq8}
\end{equation}
where $i=1, 2$ for quark and anti-quark, respectively,
 and
$J(i)=(-1)^{i+1}$.

Introducing the notations $\varphi^{\pm\pm}_{p}(q_{\perp})$ as:
\begin{equation}
\varphi^{\pm\pm}_{p}(q_{\perp})\equiv
\Lambda^{\pm}_{1p}(q_{\perp})
\frac{\not\!{P}}{M}\varphi_{p}(q_{\perp}) \frac{\not\!{P}}{M}
\Lambda^{{\pm}}_{2p}(q_{\perp})\;. \label{eq10}
\end{equation}

With contour integration over $q_{p}$ on both sides of
Eq.~(\ref{eq6}), we obtain:
$$
\varphi_{p}(q_{\perp})=\frac{
\Lambda^{+}_{1p}(q_{\perp})\eta_{p}(q_{\perp})\Lambda^{+}_{2p}(q_{\perp})}
{(M-\omega_{1}-\omega_{2})}- \frac{
\Lambda^{-}_{1p}(q_{\perp})\eta_{p}(q_{\perp})\Lambda^{-}_{2p}(q_{\perp})}
{(M+\omega_{1}+\omega_{2})}\;,
$$
and the full Salpeter equation:
$$
(M-\omega_{1}-\omega_{2})\varphi^{++}_{p}(q_{\perp})=
\Lambda^{+}_{1p}(q_{\perp})\eta_{p}(q_{\perp})\Lambda^{+}_{2p}(q_{\perp})\;,
$$
$$(M+\omega_{1}+\omega_{2})\varphi^{--}_{p}(q_{\perp})=-
\Lambda^{-}_{1p}(q_{\perp})\eta_{p}(q_{\perp})\Lambda^{-}_{2p}(q_{\perp})\;,$$
\begin{equation}
\varphi^{+-}_{p}(q_{\perp})=\varphi^{-+}_{p}(q_{\perp})=0\;.
\label{eq11}
\end{equation}

For the different $J^{PC}$ (or $J^{P}$) states, we give the general form of
wave functions. Reducing the wave functions by the last
equation of Eq.~(\ref{eq11}), then solving the first and second equations in Eq.~(\ref{eq11}) to
get the wave functions and mass spectrum. We have discussed the
solution of the Salpeter equation in detail in Ref.~\cite{w1,mass1}.

The normalization condition for BS wave function is:
\begin{equation}
\int\frac{q_{_T}^2dq_{_T}}{2{\pi}^2}Tr\left[\overline\varphi^{++}
\frac{{/}\!\!\!
{P}}{M}\varphi^{++}\frac{{/}\!\!\!{P}}{M}-\overline\varphi^{--}
\frac{{/}\!\!\! {P}}{M}\varphi^{--}\frac{{/}\!\!\!
{P}}{M}\right]=2P_{0}\;. \label{eq12}
\end{equation}

 In our model, the instantaneous interaction kernel $V$ is Cornell
potential, which is the sum of a linear scalar interaction and a vector interaction:
\begin{equation}\label{vrww}
V(r)=V_s(r)+V_0+\gamma_{_0}\otimes\gamma^0 V_v(r)= \lambda
r+V_0-\gamma_{_0}\otimes\gamma^0\frac{4}{3}\frac{\alpha_s}{r}~,
\end{equation}
 where $\lambda$ is the string constant and $\alpha_s(\vec
q)$ is the running coupling constant. In order to fit the data of
heavy quarkonia, a constant $V_0$ is often added to confine
potential. We introduce a factor $e^{-\alpha r}$ to avoid
the infrared divergence in the momentum space:
\begin{equation}
V_s(r)=\frac{\lambda}{\alpha}(1-e^{-\alpha r})~,
~~V_v(r)=-\frac{4}{3}\frac{\alpha_s}{r}e^{-\alpha r}~.
\end{equation}\label{vsvv}
 It is easy to
know that when $\alpha r\ll1$, the potential becomes to Eq.~(\ref{vrww}). In the momentum space and the C.M.S of the bound state,
the potential reads :
$$V(\vec q)=V_s(\vec q)
+\gamma_{_0}\otimes\gamma^0 V_v(\vec q)~,$$
\begin{equation}
V_s(\vec q)=-(\frac{\lambda}{\alpha}+V_0) \delta^3(\vec
q)+\frac{\lambda}{\pi^2} \frac{1}{{(\vec q}^2+{\alpha}^2)^2}~,
~~V_v(\vec q)=-\frac{2}{3{\pi}^2}\frac{\alpha_s( \vec q)}{{(\vec
q}^2+{\alpha}^2)}~,\label{eq16}
\end{equation}
where the running coupling constant $\alpha_s(\vec q)$ is :
$$\alpha_s(\vec q)=\frac{12\pi}{33-2N_f}\frac{1}
{\log (a+\frac{{\vec q}^2}{\Lambda^{2}_{QCD}})}~.$$ We introduce a small
parameter $a$ to
avoid the divergence in the denominator. The constants $\lambda$, $\alpha$, $V_0$ and
$\Lambda_{QCD}$ are the parameters that characterize the potential. $N_f=3$ for $\bar bq$ (and $\bar cq$) system.
}

\end{document}